\documentclass{article}
\usepackage{spconf,amsmath,graphicx,hyperref}
\usepackage{url}
\usepackage{booktabs}
\usepackage{amssymb}
\usepackage{xcolor}

\title{A Robust Multi-Scale Framework with Test-Time Adaptation for sEEG-Based Speech Decoding}
\name{Suli Wang\textsuperscript{1,2,\ddag},
Yang-yang Li\textsuperscript{1,3,\ddag}, 
Siqi Cai\textsuperscript{4,*}, \text{Member, IEEE},
Haizhou Li\textsuperscript{1}, \text{Fellow, IEEE}, 
\thanks{This work was supported in part by the Program for Guangdong Introducing Innovative and Entrepreneurial Teams (2023ZT10X044) and the Deutsche Forschungsgemeinschaft (DFG, EXC 2077).}}
\address{\textsuperscript{1} School of Artificial Intelligence, School of Data  Science, SRIBD, \\The Chinese University of Hong Kong, Shenzhen, China\\
\textsuperscript{2} Department of Computer Science, Technical University of Darmstadt \\
\textsuperscript{3} School of Computer Science and Engineering, Nanjing University of Science and Technology\\
\textsuperscript{4} School of Intelligence Science and Engineering, College of Artificial Intelligence, \\ Harbin Institute of Technology, Shenzhen, China
}

\begin{document}
\maketitle

\begingroup
\renewcommand\thefootnote{}
\footnotetext{\textsuperscript{\ddag}\,Equal contribution.\quad
\textsuperscript{*}Corresponding author: \href{mailto:caisiqi@ieee.org}{caisiqi@ieee.org}.}
\endgroup

\begin{abstract}
Decoding speech from stereo-electroencephalography (sEEG) signals has emerged as a promising direction for brain-computer interfaces (BCIs). Its clinical applicability, however, is limited by the inherent non-stationarity of neural signals, which causes domain shifts between training and testing, undermining decoding reliability. To address this challenge, a two-stage framework is proposed for enhanced robustness. First, a multi-scale decomposable mixing (MDM) module is introduced to model the hierarchical temporal dynamics of speech production, learning stable multi-timescale representations from sEEG signals. Second, a source-free online test-time adaptation (TTA) method performs entropy minimization to adapt the model to distribution shifts during inference. Evaluations on the public DU-IN spoken word decoding benchmark show that the approach outperforms state-of-the-art models, particularly in challenging cases. This study demonstrates that combining invariant feature learning with online adaptation is a principled strategy for developing reliable BCI systems. Source code will be released at \url{https://github.com/lyyi599/MDM-TENT}.
\end{abstract}

\begin{keywords}
Brain-Computer Interface (BCI), Speech Decoding, Stereo-Electroencephalography (sEEG), Online Decoding, Test-Time Adaptation
\end{keywords}

\section{Introduction}
The direct decoding of speech from neural activity represents a critical frontier in brain-computer interface (BCI) research, holding particular promise for individuals with amyotrophic lateral sclerosis (ALS) and other neurological conditions that result in loss of speech~\cite{metzger2023high,cho2023neural}. Among intracranial recording techniques, stereo-electroencephalography (sEEG) has emerged as a particularly promising modality for such clinical applications~\cite{he2025vocalmind}. It offers a favorable trade-off by providing high-fidelity signals from both deep and superficial cortical structures with a lower clinical risk profile compared to electrocorticography (ECoG)~\cite{miller2020current}.

Despite these advantages, the practical deployment of sEEG-based decoders is critically hindered by the inherent non-stationarity of neural signals~\cite{snider1998classification,fu2014adaptive}. The statistical properties of neural activity shift over time due to changes in cognitive state, attention, fatigue, or microscopic electrode movements~\cite{huang2024attentional, ung2017intracranial}. These variations cause a recurring out-of-distribution (OOD) generalization problem, where models trained on a subset of trials often fail to generalize to unseen trials, even within the same subject and session~\cite{saha2020intra}. Such temporal domain shifts constitute a major bottleneck for the long-term reliability and practical deployment of brain–computer interface (BCI) systems, and they account for the limited translation of high-performing laboratory models into robust clinical solutions.

\begin{figure*}
    \label{model}
    \centering
    \includegraphics[width=0.9\linewidth]{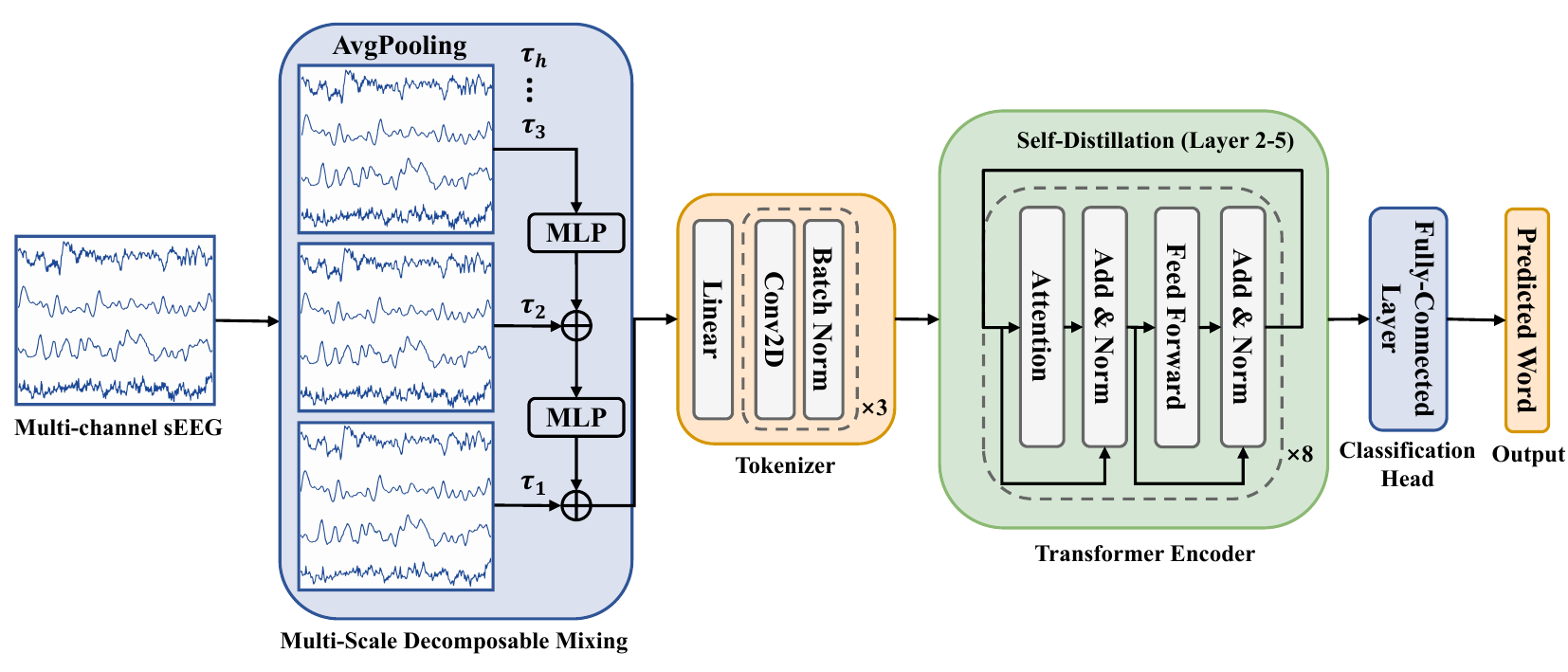}
    \caption{Base architecture. Multi-channel sEEG signals are first processed by the Multi-Scale Decomposable Mixing (MDM) module, which captures hierarchical temporal dynamics at multiple timescales. A Tokenizer then converts these features into a sequence of patch embeddings, which are subsequently passed through an 8-layer Transformer Encoder. A self-distillation scheme regularizes training and enhances feature robustness. Finally, a fully connected classification head maps the encoder’s output to the predicted word.}
    \label{fig:model}
\end{figure*}

Recent research has sought to overcome these limitations through advanced deep learning architectures. Frameworks like DU-IN have set a state-of-the-art benchmark by employing self-supervised masked modeling to learn discriminative features from sEEG data, establishing a strong baseline for the field~\cite{zheng2024discrete}. Although DU-IN provides a strong representational foundation, its architecture does not explicitly capture the multi-timescale nature of speech processing and lacks intrinsic mechanisms to address domain shift during inference.

To address these challenges, we propose a novel framework for robust sEEG-based speech decoding that integrates multi-scale representation learning with online adaptation. First, we introduce a brain-inspired Multi-Scale Decomposable Mixing (MDM) module, designed to capture hierarchical temporal dynamics. This design is motivated by neuroscientific findings which suggest that language production involves concurrent processes on multiple time scales, such as rapid motor articulation and slower lexical planning~\cite{levelt1999theory,hagoort2004integration}. Second, we employ a source-free Test-Time Adaptation (TTA) strategy based on entropy minimization~\cite{xiao2024beyond,sun2020test}. This approach directly mitigates the OOD problem during inference by enabling adaptation to the statistics of unlabeled test data without requiring access to the original training dataset. By jointly learning multi-scale representations and adapting to incoming data, our proposed framework is expected to provide a robust solution for decoding speech in dynamic, real-world environments.

\section{Method}

\subsection{Architecture Overview}

As illustrated in Fig.~\ref{model}, our proposed model is designed for speech decoding from multi-channel sEEG signals. The architecture is built upon a patch-based Transformer encoder, which processes signals through a spatial encoder (for cross-channel integration) and a temporal Transformer (for capturing sequential context)~\cite{zheng2024discrete}. To address the challenge of modeling neural activity across diverse temporal scales, we integrate a novel Multi-Scale Decomposable Mixing (MDM) module. Furthermore, to mitigate session-specific domain shift, we propose an online adaptation algorithm that updates the model during inference.

\subsection{Multi-Scale Decomposable Mixing (MDM)}

To capture the hierarchical temporal structure of neural activity underlying speech, the MDM module~\cite{horstemeyer2009multiscale} is introduced. The module constructs progressively coarser representations of the input sequence $\tau^1$ by recursively applying average pooling:

\begin{equation}
\tau^{i} = \text{AvgPooling}(\tau^{i-1}), \quad i = 2, \ldots, h,
\end{equation}
where $\tau^i$ denotes the representation at scale $i$, and $h$ is the number of hierarchical levels. This process generates a temporal pyramid, allowing information to be processed across multiple resolutions.

The multi-scale features are then fused through a residual top-down pathway. At each level, the representation is updated by integrating information from the coarser level above through a low-rank Multi-Layer Perceptron ($MLP_r$):

\begin{equation}
\xi^{i} = \tau^{i} + MLP_{r}(\xi^{i+1}), \quad \xi^{h} = \tau^{h},
\end{equation}
where $\xi^i$ denotes the fused feature at level $i$. The MLP design reduces computational overhead while retaining expressive capacity, enabling efficient processing of long sEEG sequences in practical BCI applications. Moreover, the MDM block is lightweight and preserves the input–output dimensionality, which facilitates seamless integration into diverse neural decoding frameworks. The final mixed feature $\xi^1$ provides a rich temporal abstraction that reinforces the representational capacity of the model.

\begin{table*}[!ht]
    \centering
    \caption{Performance across subjects (01--12). Results are reported as mean(\%)$\pm$std.}
    \label{performance}

    \begin{tabular}{l c c c c c c c}
    \toprule
    \multicolumn{8}{c}{\textbf{Subjects 01--06}} \\
    \midrule
    Methods & Model Size & subj-01 & subj-02 & subj-03 & subj-04 & subj-05 & subj-06 \\
    \midrule
    EEGNet~\cite{lawhern2018eegnet}        & 0.014M & 50.16$\pm$1.48 & 60.32$\pm$1.87 & 12.99$\pm$1.12 & 36.83$\pm$1.95 & 49.62$\pm$2.79 & 25.34$\pm$2.73 \\
    CNN-BiGRU~\cite{moses2021neuroprosthesis}     & 0.542M & 45.94$\pm$4.08 & 66.03$\pm$1.71 &  7.36$\pm$1.40 & 22.73$\pm$4.42 & 52.41$\pm$7.03 & 30.41$\pm$1.38 \\
    EEG-Conformer~\cite{song2022eeg} & 2.340M & 60.47$\pm$2.75 & 68.59$\pm$1.30 & 20.49$\pm$1.28 & 50.44$\pm$3.51 & 67.82$\pm$1.66 & 31.18$\pm$2.02 \\
    DU-IN~\cite{zheng2024discrete}         & 4.380M & 71.04$\pm$2.28 & 71.78$\pm$2.74 & 27.99$\pm$4.05 & 60.60$\pm$3.01 & 69.97$\pm$3.08 & 32.19$\pm$1.89 \\
    MDM-Tent (Ours)        & 5.964M & \textbf{76.24$\pm$2.62} & \textbf{76.03$\pm$1.52} & \textbf{34.63$\pm$3.81} & \textbf{71.58$\pm$1.45} & \textbf{73.88$\pm$1.95} & \textbf{37.15$\pm$1.97} \\
    \bottomrule
    \end{tabular}

    \vspace{0.6em}

    \begin{tabular}{l c c c c c c c}
    \toprule
    \multicolumn{8}{c}{\textbf{Subjects 07--12}} \\
    \midrule
    Methods & Model Size & subj-07 & subj-08 & subj-09 & subj-10 & subj-11 & subj-12 \\
    \midrule
    EEGNet~\cite{lawhern2018eegnet}        & 0.014M & 35.64$\pm$3.94 & 29.78$\pm$2.49 & 48.23$\pm$2.06 & 18.88$\pm$1.21 & 40.40$\pm$3.91 & 28.87$\pm$2.32 \\
    CNN-BiGRU~\cite{moses2021neuroprosthesis}     & 0.542M & 45.49$\pm$2.20 & 22.86$\pm$2.02 & 42.60$\pm$3.72 &  7.92$\pm$1.03 & 28.55$\pm$6.57 & 16.12$\pm$2.67 \\
    EEG-Conformer~\cite{song2022eeg} & 2.340M & 47.15$\pm$2.82 & 40.48$\pm$1.24 & 53.35$\pm$2.15 & 23.90$\pm$2.19 & 51.54$\pm$2.47 & 36.16$\pm$3.80 \\
    DU-IN~\cite{zheng2024discrete}          & 4.380M & 45.31$\pm$2.69 & 45.90$\pm$2.92 & 56.87$\pm$2.38 & 27.02$\pm$2.60 & 66.21$\pm$2.79 & 49.63$\pm$4.51 \\
    MDM-Tent (Ours)            & 5.964M & \textbf{51.19$\pm$2.98} & \textbf{51.59$\pm$1.63} & \textbf{64.53$\pm$1.95} & \textbf{37.89$\pm$1.91} & \textbf{73.25$\pm$2.73} & \textbf{61.57$\pm$4.04} \\
    \bottomrule
    \end{tabular}
\end{table*}

\subsection{Test-Time Adaptation by Entropy Minimization}

Domain shift during inference is addressed using the test-time adaptation method Tent~\cite{wang2020tent}. This approach adapts the model online using only unlabeled test data, making it particularly suitable for real-world BCI applications where recalibration is infeasible and source training data are unavailable.

The adaptation objective is the minimization of Shannon entropy for predictions on each test batch. For a prediction vector $\hat{y}$, the entropy is defined as:

\begin{equation}
    H(\hat{y}) = - \sum_{c} p(\hat{y}_{c}) \log p(\hat{y}_{c}).
\end{equation}

Lower entropy corresponds to higher model confidence and is generally associated with reduced prediction error. Entropy minimization therefore provides a self-supervised signal for adaptation.

To mitigate catastrophic forgetting, our adaptation strategy is restricted solely to the normalization layers rather than the entire network. For each test batch, this process involves:
\begin{enumerate}
    \item Estimation of Normalization Statistics: The batch-specific mean ($\mu$) and standard deviation ($\sigma$) are computed and applied to normalize features.
    \item Optimization of Affine Parameters: The affine transformation parameters ($\gamma$, $\beta$) of the normalization layers are updated using gradient descent on the entropy loss.
\end{enumerate}
This mechanism is source-free, computationally efficient, and preserves the integrity of the pre-trained backbone while enabling rapid adaptation to the statistical properties of target data, making it well-suited for online BCI systems.


\subsection{Training Objective and Self\mbox{-}Distillation}

The model is trained end-to-end for supervised word classification using cross-entropy loss. Stability and generalization are improved by applying a layer-wise self-distillation (SD) scheme~\cite{zhang2021self} as a regularizer. In this scheme, deeper representations serve as soft targets for shallower layers, promoting internal consistency and yielding more robust early-layer features.

\begin{figure*}
    \centering
    \includegraphics[width=0.85\linewidth,clip,trim={0mm 25mm 0mm 20mm}]{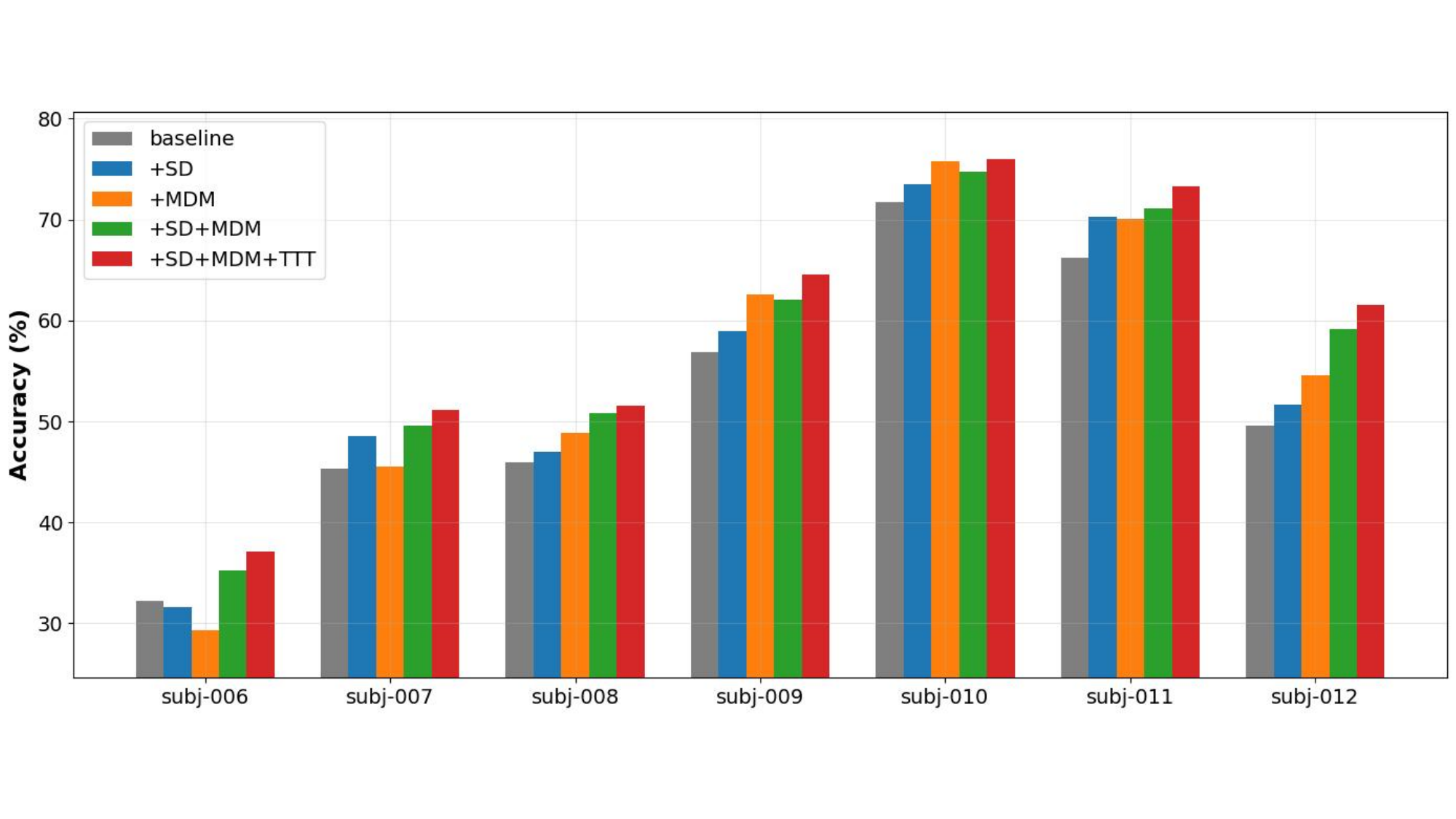}
    \caption{Ablation study on subjects 06--12. Due to space constraints, only a subset is shown. SD, MDM, and Tent were integrated sequentially; each component yielded measurable gains.}
    \label{fig:ablation}
\end{figure*}

\section{Experiments}
\label{sec:experiments}

\subsection{Dataset and Preprocessing}
Experiments were conducted on the public DU-IN dataset~\cite{zheng2024discrete}, which contains sEEG recordings from 12 subjects performing an overt speech task with 61 distinct Mandarin words. Each subject contributed approximately 3,000 trials, with each trial lasting 3 seconds (0.5s preparation, 2s articulation, and 0.5s rest). The preprocessing pipeline, including band-pass filtering, notch filtering, bipolar re-referencing, and z-score normalization, was kept consistent with the original study to ensure comparability.

Following previous studies, all evaluations were performed in a subject-dependent way. Each experiment was repeated with six different random seeds, and the mean and standard deviation of classification accuracy were reported to ensure reliability.

\subsection{Baselines} 
The proposed framework was compared with several representative neural decoding models: EEGNet~\cite{lawhern2018eegnet}, a widely used architecture for EEG decoding; CNN-BiGRU~\cite{moses2021neuroprosthesis}, a hybrid recurrent model; EEG-Conformer~\cite{song2022eeg}, a Transformer-based architecture; and the state-of-the-art DU-IN model~\cite{zheng2024discrete}, which served as the primary baseline.

\section{Results}
\subsection{Performance}
As shown in Table \ref{performance}, the proposed framework (MDM-Tent) consistently outperforms all baseline models across individual subjects and on the overall cohort. The improvement over the DU-IN baseline is particularly notable in challenging decoding scenarios. For example, on subject-03 and subject-10, where baseline accuracy is relatively low, absolute gains of 6.64\% and 10.87\% are achieved, respectively. These results highlight the robustness of the proposed approach in cases with low signal quality or pronounced domain shifts, conditions frequently encountered in clinical applications. Overall, the consistent improvements, especially in difficult cases, underscore the generalizability and effectiveness of the integrated approach.

\subsection{Ablation Study}
Ablation experiments were conducted to evaluate the contributions of individual components and their interactions. Starting from the DU-IN baseline, the training regularizer (Self-Distillation, SD), the architectural modification (MDM), and the online adaptation strategy (Tent) were added incrementally. Due to space constraints, only a subset of subjects (06--12) is shown in Fig.~\ref{fig:ablation}. Full per-subject results (01--12) are available at \url{https://github.com/lyyi599/MDM-TENT}.

The results in Fig.~\ref{fig:ablation} substantiate the effectiveness of the proposed components. Relative to the baseline (52.02\% mean accuracy), adding self-distillation (+SD) yields a modest but consistent gain to 54.03\% (+2.01\%), confirming its role as a stabilizing regularizer. Combining SD with MDM (+SD+MDM) produces a larger improvement to 57.63\% (+5.61\%), indicating complementary benefits from multi-scale feature extraction and distilled training signals. On a few subjects (e.g., subj-06), applying SD or MDM alone slightly reduces accuracy; however, their joint application recovers and surpasses the baseline, while on other subjects the improvements are more pronounced. These observations highlight the regularization effect of SD and the advantage of MDM for modeling multi-timescale dynamics.

The full configuration (+SD+MDM+Tent) achieves the highest overall accuracy, with pronounced gains on challenging subjects (subj-10 and subj-12). Leveraging the stable multi-scale backbone established by MDM and SD, Tent mitigates distributional shifts at inference and maintains robustness under clinically realistic variability.



\section{Conclusion}

This work addressed the challenge of signal non-stationarity in sEEG-based speech decoding. An integrated framework combines a neuro-inspired multi-scale architecture (MDM), a self-distillation (SD) scheme for robust feature learning, and a source-free test-time adaptation mechanism (Tent) to mitigate inference-time domain shifts. Extensive per-subject evaluations on the public DU-IN benchmark demonstrate state-of-the-art word-decoding performance, with notable gains in challenging scenarios.

Ablation analysis confirmed the synergy among components. The stable representational foundation provided by MDM and SD enables the adaptation mechanism to operate more effectively and reliably. This two-stage paradigm, in which representational capacity is first strengthened and then online adaptation is applied, offers a principled strategy for developing reliable, real-world neuroprosthetic devices.



\bibliographystyle{IEEEbib}
\bibliography{strings,refs}

\end{document}